\documentclass{aip-cp}
\usepackage[numbers]{natbib}
\usepackage{rotating}
\usepackage{graphicx}

\usepackage{verbatim}
\usepackage{savesym}
\savesymbol{iint}
\savesymbol{iiint}
\savesymbol{iiiint}
\savesymbol{idotsint}
\usepackage{amsmath}
\restoresymbol{AMS}{iint}
\restoresymbol{AMS}{iiint}
\restoresymbol{AMS}{iiiint}
\restoresymbol{AMS}{idotsint}

\graphicspath{{./figs/}}

\begin{document}

\title{Simulation of Kerr Nonlinearity: Revealing Initial State Dependency }

\author[aff1,aff2]{Souvik Agasti\corref{cor1}}

\affil[aff1]{ 	IMOMEC division, IMEC, Wetenschapspark 1, B-3590 Diepenbeek, Belgium}
\affil[aff2]{ Institute for Materials Research (IMO), Hasselt University,	Wetenschapspark 1, B-3590 Diepenbeek, Belgium}

\corresp[cor1]{ souvik.agasti@uhassest.be}

\maketitle

	\begin{abstract}
	We simulate coherent driven free dissipative Kerr nonlinear system numerically using time-evolving block decimation (TEBD) algorithm and time propagation on the Heisenberg equation of motion using Euler’s method to study how the numerical results are analogous to classical bistability
	. The system evolves through different trajectories to stabilize different branches for different external drives and initial conditions.
	The Wigner state reprentation confirms the system to suffer a residual effect of initial state throughout the non-classical dynamical evolution and the steady state of the system
	. Furthermore, 
	we also see the numerically simulated spectral density remains significantly different from analytical counterparts when initial states do not lie to the same branch of the final state.
\end{abstract}

\maketitle

\twocolumn

\section{INTRODUCTION}




The Kerr effect --i.e. non-linear quadratic electro-optic (QEO) response -- had been in interest over centuries, for having some of the interesting quantum phenomenon e.g. photon switching in quantum interference \citep{Photon_Switching}, photon bunching and antibunching in bistable steady-state \citep{Kerr_Drummond_Walls}, dynamical optical bistability via bifurcation process \citep{Yasser_Sharaby_bifurcation} and the generation of non-classical states \citep{Agasti_kerr}. In general, the multistability of nonlinear systems offers a window for a long-range of applications. Starting from first experimentally achieving a bistable state inside a Fabry-Perot  optical device cavity  \citep{gibbs_bistable_kerr}, it gives immense motivation for the construction of switching elements for potential use in optical communications and computation \citep{Photon_Switching, Optical_bistable_switch, Kerr_switch_optical_microcavity}, nanophotonic devices such as low-dimensional tunable photo-sensors \citep{kerr_photosensor},  magneto-optical storage devices \citep{magneto_optical_storage_devices}, NRZ-to-RZ conversion \cite{NRZ2RZ}, regeneration, monitoring, multi-casting, demultiplexing and multiple-wavelength generation \cite{multiple-wavelength_source_kerr1, multiple-wavelength_source_kerr2}. 
This prospect of technological applications demands to have sufficient knowledge of their frequency response, relative phase, and squeezed phase parameters to engineer the performance of the devices and control and manipulation of signals.

In this context, it has been noticed that, with the influence of an appropriate external field, the non-linear behavior of the system generates a specific quantum state in the system. The Kerr non-linearity itself is a prominent example which generates two-mode squeezed vacuum (TMSV) by spontaneous parametric down-conversion process \citep{Boyd_Nonlinear_Optics}.  Apart from Kerr non-linear systems, other examples could be optomechanical systems  where ground-state cooling \citep{optomechanics_cooling}, along with squeezing below the standard quantum limit (SQL) \citep{optomechanics_SQL} 
and quantum-limited amplification \citep{optomechanics_amplification} has been achieved by controlling the external drive. Another important example can be considered as the impurity-infected solid-state systems where the quadrature fluctuation embedded with the state of the system is not only dependent on the external drive but also the role of the initial state of the impurities \citep{myprevious_prl}. Also, the atomic system has remained able to control phase for coherent pulse propagation and switching \cite{Phase_control_switch}. Therefore, it has been of huge attention to investigate how nonlinearity and the external drive play a significant role in the control and manipulation of the system, and the impact of the initial state in the dynamical behavior of quantum systems surviving under spontaneous decoherence.

Kerr nonlinearity ensures to display bistability in all setups as a predominant characteristic at stationary state, which has been encountered in two different ways: semi-classically by approximating the state of the system to the nearest coherent state, and quantum mechanically which estimates the exact solution by using the master equation formalism of the density matrix of the system. The semi-classical solution of both dispersive and absorptive bistability has been obtained through the {Heisenberg-Langevin} equation \cite{McCall1974kerr, Agrawal_Carmichael_kerr, reviewer1_suggestion}, and the theory of quantum mechanical solution for the absorptive case \cite{Walls_Carmichael_kerr1977} and the dissipative case \cite{Kerr_Drummond_Walls} has been derived by mapping the master equation to the Fokker Planck equation. 
While implementing both the theoretical techniques on Kerr nonlinear systems, they moreover determine Markovian dynamics of open quantum systems which are considered as one of the most fundamental problems in quantum mechanics.

The open quantum dynamics has been well regarded for establishing the conceptual background of fundamental physics, encompassing concepts such as distinguishing boundary between classical and quantum physics \citep{boundary_quantum_classical} and highlighting issues in the detection of gravitational waves \cite{gravitational_wave_mypaper}. 	
However, while investigating the decoherence dynamics of nonlinear open quantum systems, the theory has been implemented after transforming the nonlinear Hamiltonian to a linear one by linearizing the quantum fluctuation over nonlinear steady-state field amplitude. {Appreciating its simplicity, this model, therefore, had remained unable to provide a satisfactory framework while obtaining the exact dynamical behavior, which leads to overlooking interesting effects introduced by non-linearity. Especially, the exact dynamical behavior of the Kerr nonlinear systems and the impact of the initial state can not be determined analytically, which requires numerical treatment to deal with. This limitation of analytical treatment explicitly motivates us to simulate the time evolution numerically}, which consists of transforming the environmental degrees of freedom to a one-dimensional (1D) many-body chain \cite{Agasti_TEBD_conference} and simulating the chain afterward. The computational method is composed of numerical diagonalization and renormalization process  \citep{Agasti_kerr, Agasti_OQS}.

The time-adaptive density matrix renormalization group (t-DMRG) algorithm is considered one of the most powerful methods in optical, atomic, and condensed matter physics to simulate strongly-correlated many-body quantum systems. 
While implementing the algorithm for open quantum systems, we map the canonical S/B coupling model to a 1D harmonic chain with nearest-neighbor interactions. Recently, such mappings have been used for the simulation of open quantum systems \cite{openquanta_1dchain} to simulate spin-boson models \cite{openquanta_1dchain_spin_bosson} and biomolecular systems \cite{openquanta_1dchain_biomolecul}.


In this article, besides analytical theory, using two different numerical techniques, i.e. Euler's time propagation, and time-evolving block decimation (TEBD) methods, we simulate Kerr non-linear system, to investigate the impact of the initial state
, under the influence of the different driving fields. Earlier, in Ref. \citep{Agasti_kerr}, while simulating the dynamics of the Kerr nonlinear system using TEBD, we have shown the consistency between the analytical and the numerical results. Also, we have observed that the TEBD numerical result follows the	quantum mechanical exact solution, whereas the time propagation of the system field obtained using Euler’s method follows the semi-classical solution of the Heisenberg equation of motion. Therefore, in this case, we especially focus on how the nature of the quantum jump is influenced by the initial state of the system. Here, we start with a brief description of the system. Following by, we show how the bistable nature of the steady-state of the system is dependent on the initial state of the system. Characterizing the effects of the non-linearity on the dynamical behavior of the system, we afterward determine the fluctuation spectra for the different initial states. Our analysis, moreover, provides a platform to look after its frequency response which will be useful in the fabrication of switching devices. The impact of the initial state on a coherent driven nonlinear system also implicates 
some of the recently observed phenomena in the context of impurity infected solid-state systems \citep{myprevious_prl}.

\begin{figure*}[t!]
	\includegraphics[width=1\linewidth]{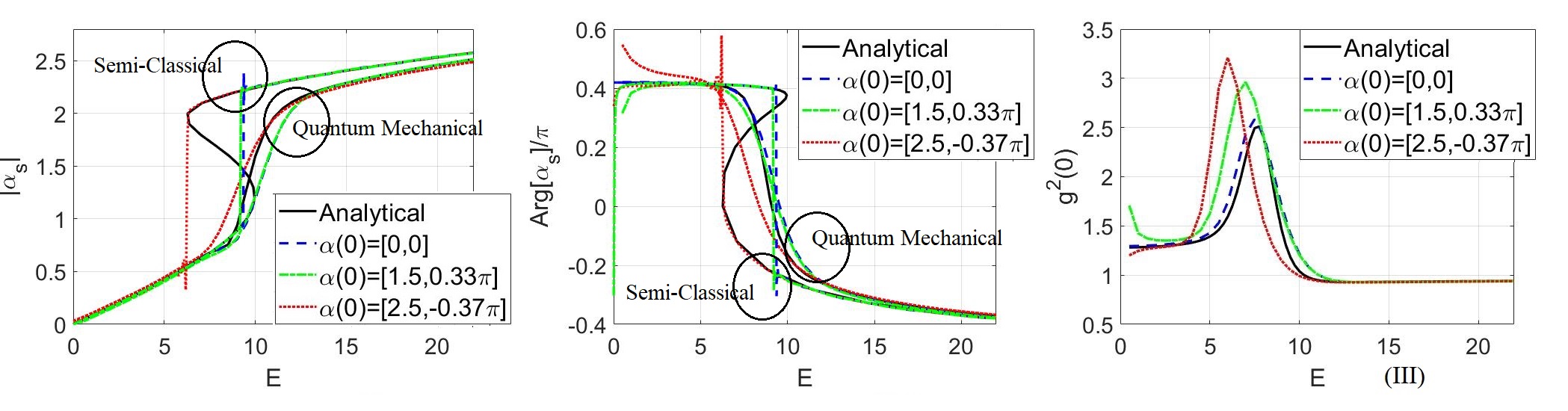}
	\caption{Steady state field amplitude and correlation function plotted with the variation of driving field amplitude for $\Delta=-12 g,\chi"= 1.5 g, \gamma= 6.28 g$. TEBD simulation parameters $N = 61, x_{max} = 60, \chi = 36, M = 20, \delta t = 10^{-2} g^{-1}$ and total time of evolution $2 g^{-1}$. }
	\label{fig:kerr_plot}
\end{figure*}

\begin{figure*}[t!]
	\includegraphics[width=1\linewidth]{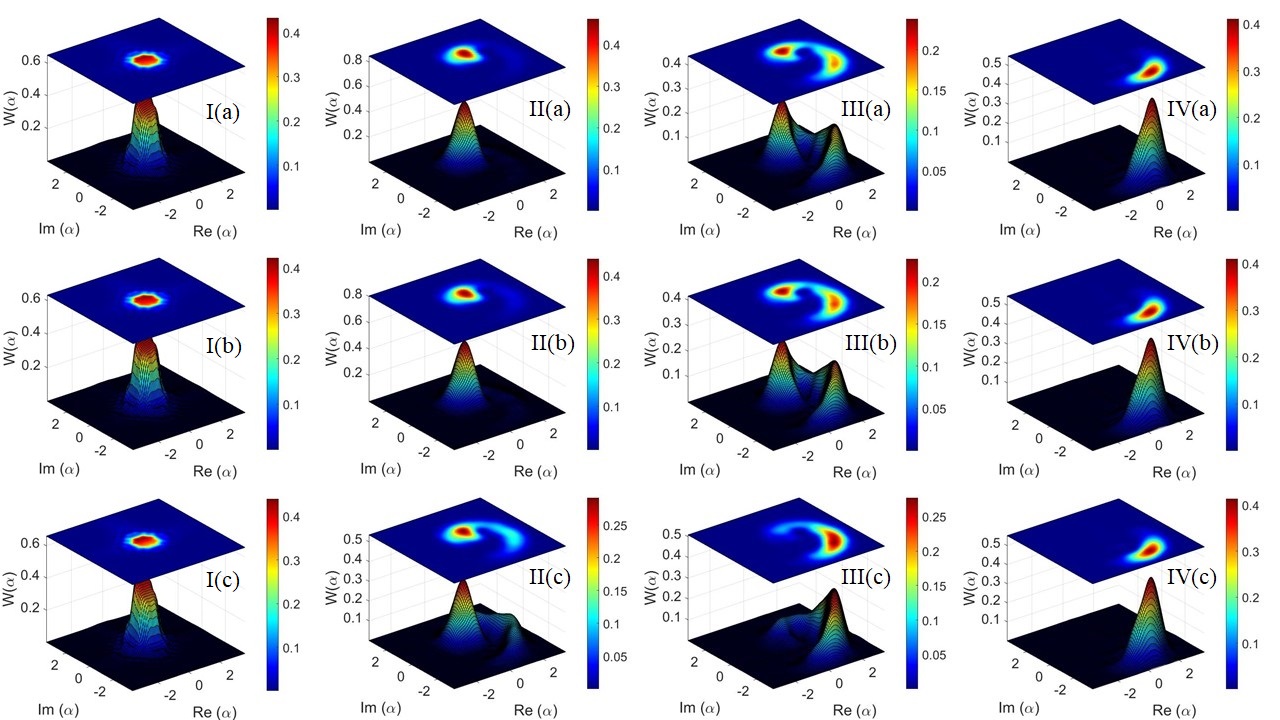}
	\caption{Steady state Wigner function for (I) E$= 1$, (II) E$=8$, (III) E$=10$ and (IV) E$= 20$, and (a)$\alpha(0)=[0,0]$, (b)$\alpha(0)=[1.5,0.33\pi]$ and (c)$\alpha(0)=[2.5,-0.37\pi]$ 
		. All other parameters remain same with Fig. \ref{fig:kerr_plot}.}
	\label{fig:wig_fn,E=1,8,10,20(sp0_1,5_2,5)}
\end{figure*}

\begin{figure*}[t!]
	\includegraphics[width=1\linewidth]{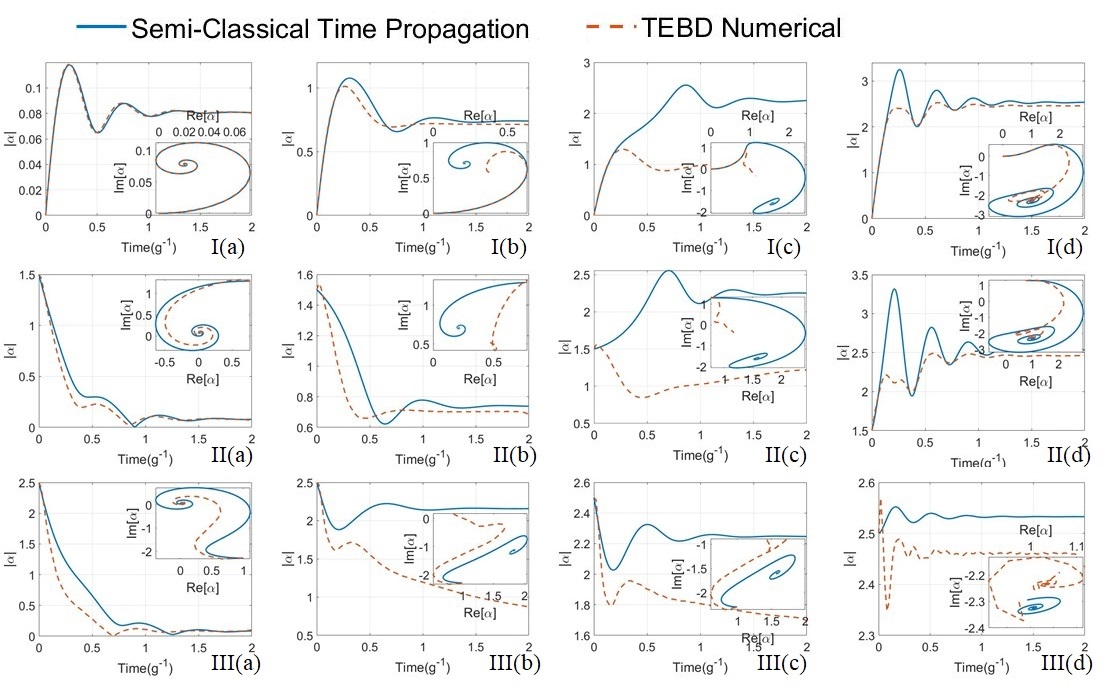}
	\caption{Dynamical behaviour of the system field with the variation of driving field amplitude for (a) E$= 1$, (b) E$=8$, (c) E$=10$ and (d) E$= 20$, and initial field (I)$\alpha(0)=[0,0]$, (II)$\alpha(0)=[1.5,0.33\pi]$ and (III)$\alpha(0)=[2.5,-0.37\pi]$. All other parameters remain same with Fig. \ref{fig:kerr_plot}. In inset we plot the trajectory of the system field, along with the phase. }
	\label{fig:kerr_dyne}
\end{figure*}

\begin{figure*}[t!]
	\centering
	\includegraphics[width=1\linewidth]{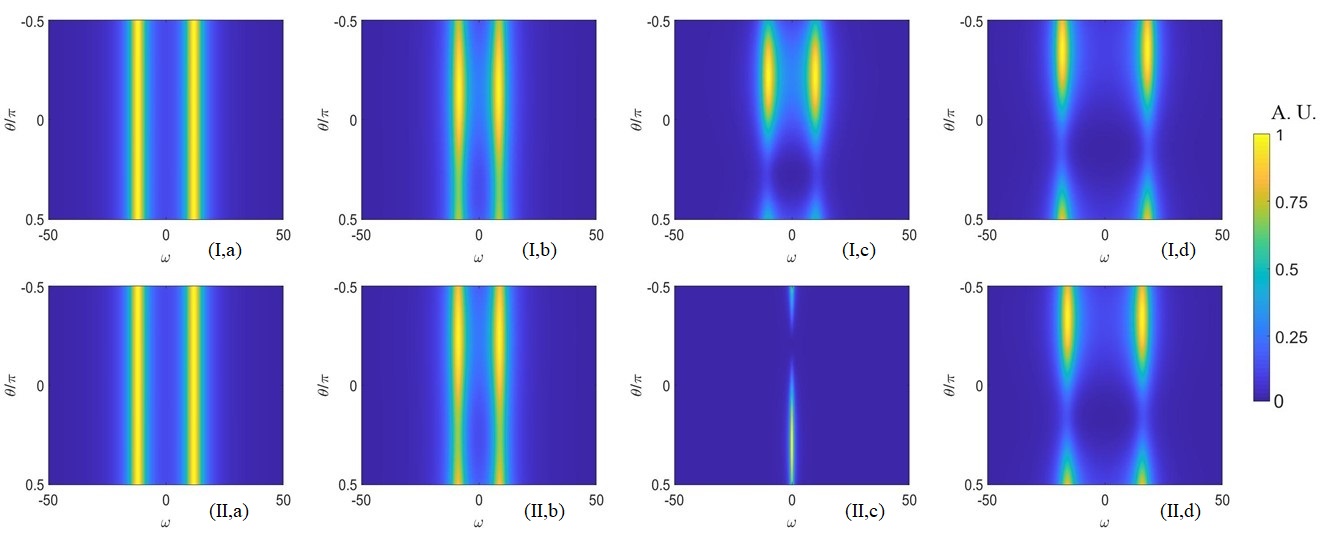}
	\caption{Analytically determined fluctuation spectra for (a) E$= 1$, (b) E$=8$, (c) E$=10$ and (d) E$= 20$ determined using (I) semi-classical and  (II) quantum mechanical treatments. All other parameters remain same with Fig. \ref{fig:kerr_plot}. }
	\label{anal_fluc(E1,8,10,20)}
\end{figure*}

\begin{figure*}[t!]
	\centering
	\includegraphics[width=1\linewidth]{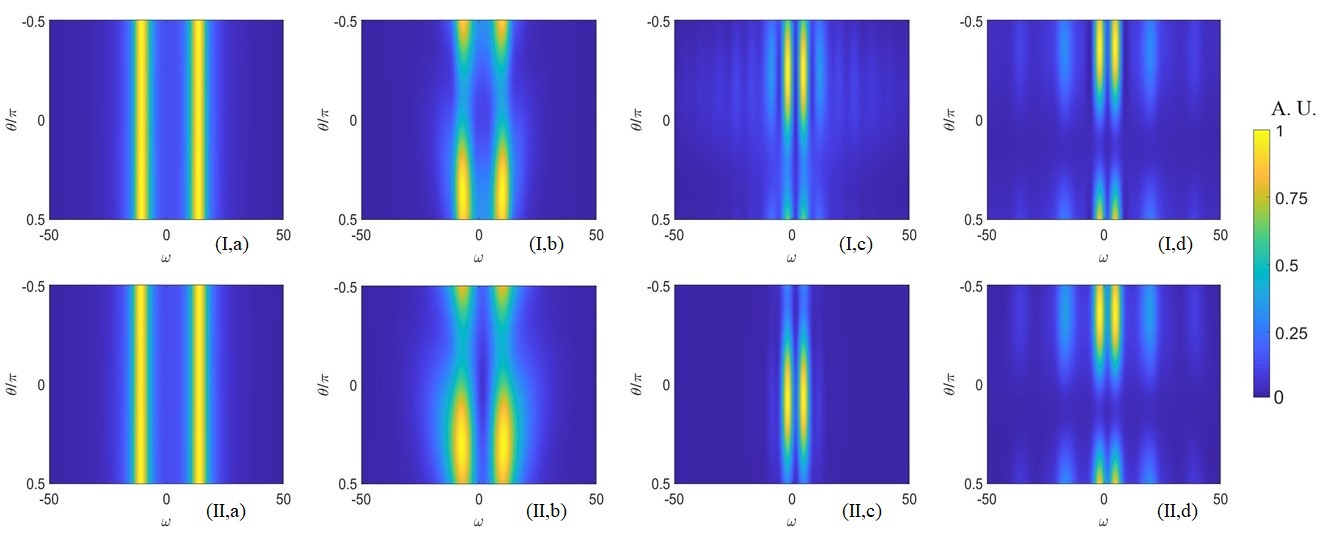}
	\caption{ Numerically determined fluctuation spectra for (a) E$= 1$, (b) E$=8$, (c) E$=10$ and (d) E$= 20$ determined using (I) Euler’s time propagation and  (II) TEBD numerical methods, initial field $\alpha(0)=0$. All other parameters remain same with Fig. \ref{fig:kerr_plot}. }
	\label{numr_fluc(E1,8,10,20)(sp0)}
\end{figure*}

\begin{figure*}[t]
	\centering
	\includegraphics[width=1\linewidth]{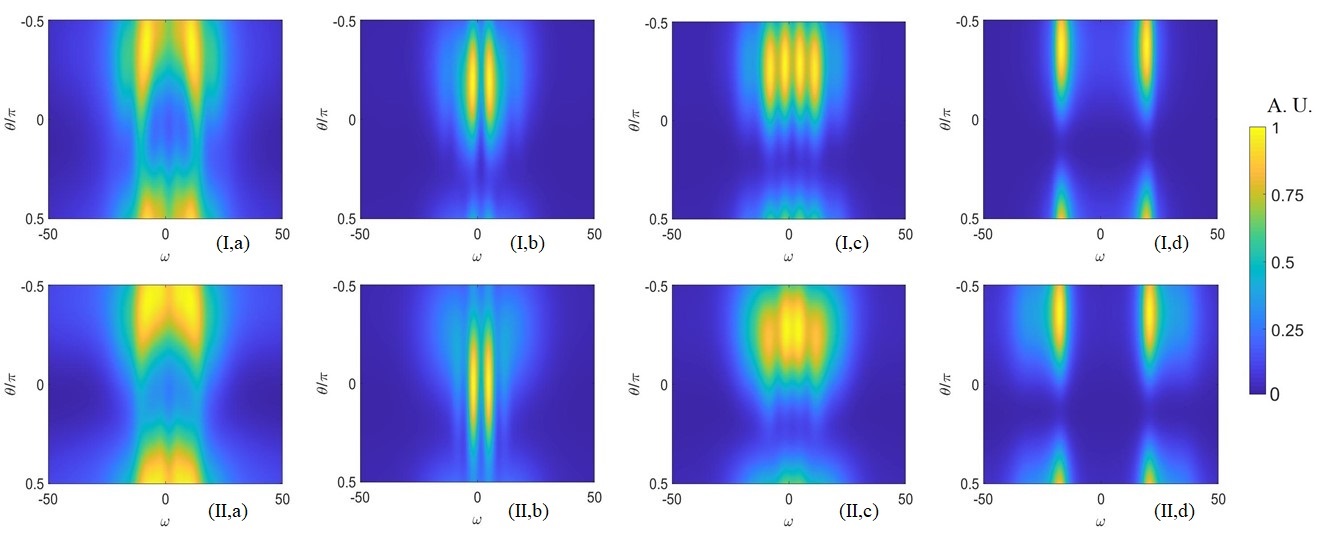}
	\caption{Numerically determined fluctuation spectra for (a) E$= 1$, (b) E$=8$, (c) E$=10$ and (d) E$= 20$ determined using (I) Euler’s time propagation and  (II) TEBD numerical methods, initial field $\alpha(0) =[2.5,-0.37\pi]$. All other parameters remain same with Fig. \ref{fig:kerr_plot}. }
	\label{numr_fluc(E1,8,10,20)(sp2,5)}
\end{figure*}

\begin{figure*}[t]
	\centering
	\includegraphics[width=1\linewidth]{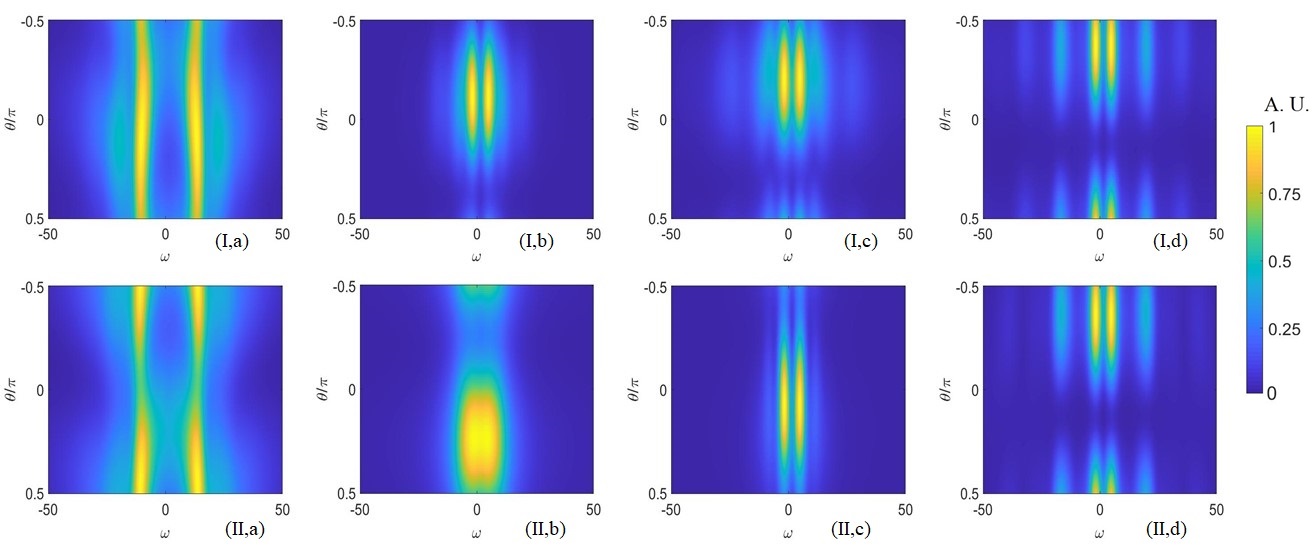}
	\caption{Numerically determined fluctuation spectra for (a) E$= 1$, (b) E$=8$, (c) E$=10$ and (d) E$= 20$ determined using (I) Euler’s time propagation and  (II) TEBD numerical methods, initial field $\alpha(0)=[1.5,0.33\pi]$. All other parameters remain same with Fig. \ref{fig:kerr_plot}. }
	\label{numr_fluc(E1,8,10,20)(sp1,5)}
\end{figure*}

\section{THEORETICAL MODEL}\label{sec:level2}

We start with the Hamiltonian of a Kerr nonlinear system, described as

\begin{equation}\label{kerr_hamiltonian}
	{H}_S =\omega_S a^\dagger a+\chi"{a^\dagger}^2a^2+i(a^\dagger Ee^{-{i\omega _{L}t}}- a E^*e^{{i\omega _{L}t}})
\end{equation}

where $\omega_S$ is the frequency of the mode of oscillation, $a^\dagger (a)$ are the creation (annihilation) operators of the system, and $\chi"$ is the anharmonicity parameter which is contributed by the real part of the third-order nonlinear susceptibility tensor
.  $\vec{\tilde{E}}(t)=\vec{E}e^{-i\omega_L t}+\vec{E}^* e^{i\omega_L t}$ is an external driving field of amplitude  $E$ and oscillation frequency $\omega_L$, applied on the system. To make the Hamiltonian time-independent, shifting to the frame of the driving field gives the detuned cavity frequency $\Delta=\omega_S-\omega_L$.
The total Hamiltonian of the  system (S) coupled to a bosonic reservoir (B) is 

\begin{equation}
	H_{tot} = H_S + H_B + H_{SB} 
\end{equation} 

where $H_B =  \lim_{x_{m}\to \infty} \int_{-x_{m}}^{x_{m}} g(x) d^\dagger (x) d(x) \mathrm{d}x $ represents the Hamiltonian of a multimode reservoir, and the interaction Hamiltonian is $H_{SB} =  \lim_{x_{m}\to \infty} \int_{-x_{m}}^{x_{m}} h(x)\left(a^\dagger d(x) + h.c.\right) \mathrm{d}x $. The thermal reservoir is at {zero temperature}. $d^\dagger_x (d_x)$  are the creation (annihilation) operators, and  $g(x)$ and $h(x)$ are the frequency of oscillation and the coupling strength between the system and environment, respectively, for the environmental mode $x$, around the central frequency $\omega_S$.
The properties of bath can be characterized by a uniquely defined spectral density function $J(\omega)$.  Considering linear dispersion relation: $g(x) = g. x,$ where $g$ is the inverse of density of states, we choose a hard cutoff limit of the frequency of the bath: $\omega_c = g.x_{m}$.
In addition, for having Markovian dynamics, within the range of frequencies of interest, the S/B coupling strength is considered to be mode independent (wide band limit approximation):  $h(x) =c_0 $ \cite{book_robert}, which gives the spectral density function  \cite{spectral_density_function} 

\begin{equation}
	\label{spctral_density}
	J(\omega) = \frac{1}{2} \gamma\Theta(\omega+\omega_c)\Theta(\omega_c-\omega),
\end{equation} 

where $\gamma = 2\pi c_0^2$ is the rate of dissipation of the system and $\Theta$ is the Heaviside step function. 
The equation of motion of the system field operators can be obtained semi-classically from the {Heisenberg-Langevin} equation of motion

\begin{align}\label{QLEfrKerr}
	\dot{a}&=-i\Delta a-2i\chi'' a^\dagger a^2 -E-\frac{\gamma}{2}a    
\end{align}

Based on the theoretical framework defined here, we determine the dynamical behavior of the Kerr nonlinear system numerically using two different methods. The first method includes the transformation of the S/B coupling method to 1D chain and simulates afterward using the TEBD algorithm { (see Appendix \ref{TEBD_NUMERICAL_MODEL}) }. The applicability and limitations of the TEBD technique are explained in \citep{Agasti_kerr} where we used the technique successfully for the first time for the simulation of the Kerr nonlinear system. The second approach involves time propagation of the semi-classical nonlinear differential equations given in the Eq. \eqref{QLEfrKerr} using Euler's method \cite{Agasti_kerr}.

The bistable nature of the Kerr nonlinear system provokes us to consider three different initial conditions, to investigate the switching effects on the dynamical behavior. Earlier, we had seen that the initial state of the impurities present in a bosonic solid-state system had a significant impact on the dynamics of the system \citep{myprevious_prl}.

\section{ STEADY STATE CONDITION} 

The exact quantum mechanical solution of the moment calculating generalized function of the system field operators is derived in Schr\"odinger picture by mapping the master equation to the Fokker-Planck equation \cite{Kerr_Drummond_Walls}, which in turn determines the steady-state field amplitude and second order correlation functions:

\begin{subequations}
	\label{Kerr master g20 and field}
	\begin{align}
		\langle a\rangle &=\left(-\frac{[E_0/i\chi] F\big(p+1,q,2[E_0/\chi]^2\big)}{pF\big(p,q,2[E_0/\chi]^2\big)}\right). \label{Kerr master field} \\
		g^2(0) &=\left(\frac{pq F\big(p,q,2|E_0/\chi|^2\big)F\big(p+2,q+2,2[E_0/\chi]^2\big)}{(p+1)(q+1)[F\big(p+1,q+1,2[E_0/\chi]^2\big)]^2}\right),\label{subeq:1}
	\end{align}
\end{subequations}

where $E_0=E=E^*, p=[\frac{\Delta}{\chi"}+\frac{\gamma}{2i\chi"}],q=[\frac{\Delta}{\chi"}-\frac{\gamma}{2i\chi"}]$ and $F\big(p,q,z\big)=F\big([],[p,q],z\big)$ is the $_0F_2$ hypergeometric function.

{
Fig. \ref{fig:kerr_plot} presents a comparison of analytically determined semi-classical (obtained from Eq. \eqref{QLEfrKerr}) { (See Appendix \ref{SEMI_CLASSICAL_EQUATION})} and quantum mechanical steady-state behavior (Eq. \eqref{Kerr master field}) of the system to their corresponding numerical counterparts which are Euler's time propagation and the TEBD numerical methods
. The amplitudes and the phases of the steady-state system fields are plotted in Fig. \ref{fig:kerr_plot} (I) and (II), respectively; which shows how the TEBD numerical result remains analogous to classical bistability. Following the semi-classical analytical solution which determines the brunch values and the transition region, the steady-state system field loses its linear behavior when we increase the external drive, and the system jumps from one steady-state to another. 
The numerical technique involved with Euler's time propagation method follows the semi-classical analytical solution, whereas TEBD  numerical technique follows the quantum mechanical analytical solution, which is also seen before in Ref. \citep{Agasti_kerr}.} Even though the semi-classical solution exhibits bistability, the quantum mechanical solution does not; rather creating a superposition of them. As the coherent states are non-orthogonal, the superposition of two coherent states in the quantum mechanical solution creates a non-classical state. As implication, we observe a peak in the plot of second-order correlation function $(g^2(0))$ in Fig. \ref{fig:kerr_plot} (III) around the transition region. The plot of $g^2(0)$ shows that the system undergoes from bunching to the anti-bunching steady-state mode when it shifts from lower brunch to upper under the influence of stronger drive. The non-classical nature of the steady-state due to the superposition of two coherent states, and the transition of the system from bunching to anti-bunching mode with the increment of an external drive is better understood from the movement of the bump of Wigner function. While increasing the strength of the drive at Fig.  \ref{fig:wig_fn,E=1,8,10,20(sp0_1,5_2,5)} (I-IV), the bump is seen to be stabilized at different locations at phase space, which clearly indicates superposition of two stable coherent states.

More interesting features are depicted when we look after the impact of the initial state on the steady-state behavior of the system. {The initial states are considered to be coherent and chosen to live in different branches, are noted by $\alpha(0)$.} Fig. \ref{fig:kerr_plot} (I) and (II) also indicates that the semi-classical jump occurs at the different driving field within the classically determined transition region, for different initial states. A residual effect is observed in the TEBD determined quantum mechanical estimation, which intends to transit from bunching to anti-bunching mode earlier for the initial state belonging in upper brunch, resulting in shifting the pick of $(g^2(0))$ in Fig. \ref{fig:kerr_plot} (III), towards the weaker drive. The phenomenon is even better visualized in the plot of Wigner function (Fig.  \ref{fig:wig_fn,E=1,8,10,20(sp0_1,5_2,5)} (a-c)), where the shift of the bump occurs earlier when the system starts evolving from an initial state belonging to the upper branch. 

Note, importantly, that the analytically determined steady-state is the result obtained ideally at infinite time after all possible transitions that can occur, which is 
independent of the initial state of the system. However, in reality, any switching device does not evolve for infinite time, rather a long time ($>> t_{damp}$) evolution that ensures the system be stabilized, is typically accepted as the final state of the system. Therefore, one must take into account the residual effects as the impact of the initial state while studying the dynamical behavior or using the Kerr effect in the design of switching systems. Continuation of the impact of the initial state also indicates that the rate of oscillations of the system field becomes extremely low around the transition region, which is also revealed from the study of the fluctuation spectrum in the following section. 
In this context, recently, in Ref. \citep{myprevious_prl} it has been noticed that for a different type of non-linearity, where the steady-state system field and the fluctuation spectra of an impurity infected bosonic solid-state system also remains distinctly dependent on the initial states of the impurities.

\section{FIELD DYNAMICS} 

Following the footsteps of privious simulation of Kerr nonlinear systems \cite{Agasti_kerr}, here, we study the impact of initial state on it's dynamical behavior. Fig. \ref{fig:kerr_dyne} shows how the system field stabilizes after initial fluctuation. Not surprisingly, {the plots exhibit difference in two different methods due to the fact that the Euler's time propagation determines the classical field of the system which lies on one among two branches, whereas TEBD generates superposition of them. As consequence, the difference enhances around the transition region}. Interestingly, the dynamical behavior is seen to be different for different initial conditions. More interesting phenomenon is noticed when we plot the trajectory of the time evolution in the phase space. The plots given in insets exhibit that the system evolves in a (counter) clockwise trajectory when it evolves in (lower) upper branch. The system tends to change its trajectory immediately when it jumps from one branch to another during the time evolution (Fig. \ref{fig:kerr_dyne}I(c,d) and III(a)). However, the phase space plot of trajectories determined by two different methods, remain significantly different, especially in the transition region (Fig. \ref{fig:kerr_dyne}(I-III)(b,c)). In fact, due to the superposition of two coherent states, no particular pattern is found in TEBD method when the system evolves majorly through transition region (Fig. \ref{fig:kerr_dyne}(I-III)(c)). {The difference in their dynamical behavior also brings a significant distingtion in the bahavior of their spectral density, which is the following topic of discussion.} 

\section{FLUCTUATION SPECTRA: ANALYTICALLY DETERMINED}

The numerical method, therefore, provides a solid platform to deal with nonlinearities to study the dynamics of Kerr nonlinear systems. To  characterize  the nonlinear effects, we evaluate the fluctuation spectrum of the field of the system
$S_\omega^{\theta}=1/2 \langle\left\{X^\theta_\omega,
X^\theta_{-\omega}\right\}\rangle$ -- with
$X_\omega^\theta = 1/\sqrt{2} \left(a_{-\omega}^\dagger e^{i  \theta}+a_\omega
e^{-i \theta}\right)$ -- taking vacuum fluctuations into account of the multimode bath. The presence of nonlinear terms in the system Hamiltonian induces squeezing, which can be experimentally observed by homodyne detection of the output field.
The analytically determined fluctuation spectrum is shown in the Fig. \ref{anal_fluc(E1,8,10,20)}
where it exhibits a clear dependence
on the frequency and phase.

\subsection{Frequency dependence on driving field}

Fig. \ref{anal_fluc(E1,8,10,20)} shows that, when the system saturates to the lower branch, the normal mode splitting of the fluctuation spectra reduces while increasing the strength of the driving field. However, an opposite phenomenon is observed when the system saturates to the upper branch, i.e. increment of the normal mode splitting while increasing the strength of the pump. The phenomenon is hinted at while investigating the system field, i.e. the rate of change becomes very low around the transition region, which is better understood from the linearized response of the fluctuations of the system field given in {Appendix} \ref{LINEARIZED_MODEL} (Eq. \eqref{split freq}).


\subsection{Phase dependence on driving field}


Fig. \ref{anal_fluc(E1,8,10,20)} also shows that the squeezing spectrum becomes phase independent for a weaker drive, which can be hinted by the fact that the impact of nonlinearity disappears and the system responses linear for a weaker drive in the lower branch (I,a and II,a).
Furthermore, the phase dependency becomes prominent with the increment of the driving field. We also notice that the fluctuation spectra determined by two different methods exhibit difference especially in transition region Fig. \ref{anal_fluc(E1,8,10,20)}((I,b), (II,b) and (I,c) (II,c)), which appears due to the difference in the amplitudes of the steady-state system field. 
However, when we go further from the transition region, as the system collapses to one of the two classical stable states the difference in the fluctuation spectra reduces. 

\section{FLUCTUATION SPECTRA: NUMERICALLY DETERMINED }

Even though the initial state of the Kerr nonlinear system has an impact on the steady-state behavior of the system, the impact is more severe on its dynamical behavior, which we investigate by determining the fluctuation spectra and comparing to its analytical counterpart. As anticipated, we see the numerical spectra determined using Euler’s time propagation and TEBD numerical methods (Figs. \ref{numr_fluc(E1,8,10,20)(sp0)}, \ref{numr_fluc(E1,8,10,20)(sp2,5)}, \ref{numr_fluc(E1,8,10,20)(sp1,5)}) exhibit similarity to their corresponding semi-classically and quantum mechanically determined spectra (Fig. \ref{anal_fluc(E1,8,10,20)}), respectively. However, the difference in the initial state exhibits a significant difference in their fluctuation spectra.


\subsection{Initial state belongs to lower brunch}

We start with ground state $(\alpha(0)=0)$, to ensure the initial state belonging to the lower branch. By plotting the fluctuation spectra in Fig. \ref{numr_fluc(E1,8,10,20)(sp0)} we see that, in case of weaker driving field when the system stabilizes to lower branch, the obtained numerical spectra remains  indifferent to the analytical estimation (Fig. \ref{anal_fluc(E1,8,10,20)} (I,a) and (II,a) ). However, a similar comparison in the higher driving field exhibits a significant difference. In particular, 
we find an intense response at comparatively lower frequency in both the numerical spectra when the system saturates to upper branch  (Fig.  \ref{numr_fluc(E1,8,10,20)(sp0)} (I,d and II,d) ), which occurs due to the transition of the system from lower branch to upper.

\subsection{Initial state belongs to upper brunch}

Hereafter, we investigate for the system belonging initially to an upper stable branch ($\alpha=[2.5,-0.37\pi]$) in Fig. \ref{numr_fluc(E1,8,10,20)(sp2,5)}, which shows that, unlike previous case, both the fluctuation spectra determined numerically moreover exhibit similar patterns to the analytical spectra (Fig. \ref{anal_fluc(E1,8,10,20)}) when the driving field is stronger (I,d and II,d). Anticipating that, however, due to the oscillation of the system while and before transiting from upper branch to lower, in the case of the lower driving field, we see multiple responses in the fluctuation spectra (Fig. \ref{numr_fluc(E1,8,10,20)(sp2,5)}(I and II a,b,c) ), exhibiting significant difference from their corresponding analytical counterpart (Fig. \ref{anal_fluc(E1,8,10,20)}).

\subsection{Initial state belongs to metastable brunch}

Finally, for the initial coherent state which belongs to the metastable brunch of the classically defined transition region ($\alpha=[1.5,0.33\pi]$), the fluctuation spectra determined numerically is plotted in Fig. \ref{numr_fluc(E1,8,10,20)(sp1,5)}, which shows that they differ from their corresponding analytical counterpart (Fig. \ref{anal_fluc(E1,8,10,20)}) significantly, not only for the cases where the system reaches to steady-state but also for the transition region, since, for TEBD numerical simulation even though the system starts evolving form an initial classical state, it ends up evolving to a non-classical state. However, for semi-classical Euler's time propagation, the final state collapses to one among stable branches (upper or lower, but not metastable), even though the evolution started from the metastable branch.

\section{CONCLUSION}\label{sec:level5}

We have used Euler’s time propagation and TEBD numerical techniques successfully to study the dynamical behavior of the Kerr nonlinear system. Unlike the conventional theoretical approach that considers linearise approximation, the method exhibits better accuracy while determining the quantum fluctuations. Determining steady-state system field, we see that, 
the time propagation of the system field obtained using Euler’s method follows the semi-classical solution of the Heisenberg equation of motion, whereas the TEBD numerical simulation follows the quantum mechanical exact solution obtained by mapping the master equation to a Fokker-Planck equation. Therefore, the semi-classical jump determined by Euler’s method for different initial states occurs at the different driving fields within the classically determined transition region. Moreover, the semi-classical Euler’s method determines a coherent field of the system which belongs to one among two branches, whereas TEBD numerical result determines the superposition of them, generating non-classical states. The phenomenon has been confirmed when we determine the second-order correlation function and Wigner function, which also reveals that the system suffers a residual effect of the initial state, as we see that, with the increment of the driving field, the system tends to jump earlier from bunching to anti-bunching mode when it starts evolving from an initial state belonging to the upper branch. The non-linearity introduced frequency and phase dependency of the noise fluctuation spectra also exhibits initial state dependency. The numerical result remains significantly different from its corresponding analytical counterpart especially when the initial state does not belong to the same branch of the final state.

Our work exhibits importance as it is capable of determining the dynamical behavior of the externally driven Kerr nonlinear system, which has been analyzed in recent experiments. For example, the influence of different magnetic fields on electrical conductivity in nonlinear media has drawn attention for exhibiting interesting quantum effects \cite{Garcia-Merino:16, Garc_a_Merino_2017, reviewer2_suggestion1, reviewer2_suggestion2}.  Besides, {it will also be beneficiary for the phase control of switching systems} \cite{Phase_control_switch}. While using for switching purposes, the nonclassical behavior around the transition region determines the acceptable range of the control drive.	
The results could be useful in the study of optical pulse propagation in nonlinear media, espicially, how the nonolinearity effects non dephasing of the optical signal \cite{optical_prop_Kerr_fiber}.
 More importantly, it is important to have an idea of their frequency response to engineer novel schemes for sensing, control, and manipulation of signals, and our analysis will be extremely useful to fulfill that goal as it determines speed while designing and manufacturing ultrafast devices such as photo-sensors \citep{kerr_photosensor}, storage systems \citep{magneto_optical_storage_devices} and switching systems \citep{Photon_Switching, Optical_bistable_switch, Kerr_switch_optical_microcavity}. Based on the performance of the numerical techniques, we conclude by saying that one can consider it to be a promising platform to handle nonlinear systems, e.g. microwave quantum optomechanics, two-level systems, and solid-state open quantum systems reported in \citep{myprevious_prl}.

\section{ACKNOWLEDGMENTS}
	SA wish to acknowledge the support of
	Muhammad Asjad, Subrata Chakraborty, Philippe Djorw\'e and Abhishek Shukla for their technical advice while working on this project.  The work has been supported by European Union; Project number: 101065991 (acronym: SingletSQL)


\nocite{*}
\bibliographystyle{aipnum-cp}%
\bibliography{sample}%

\onecolumn

\appendix

\section{APPENDIX: TEBD NUMERICAL MODEL}  \label{TEBD_NUMERICAL_MODEL}

\begin{figure*}
	\centering
	\includegraphics[width=\linewidth]{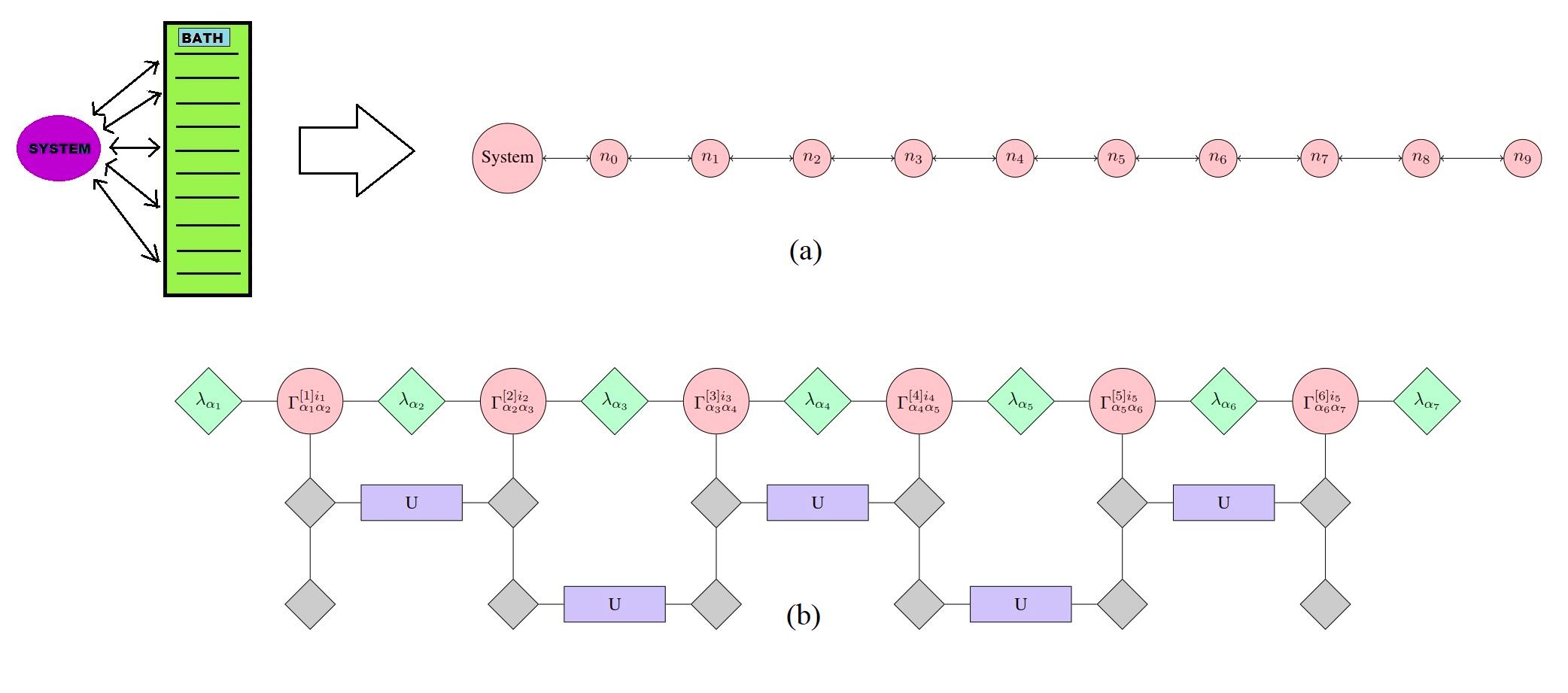}
	\caption{(a) Transformation of Hamiltonian from system/bath coupling model to semi infinite chain model. (b) Diagrammatic expression of the real time evolution operation on alternating pair}
	\label{fig:tebd}
\end{figure*}

In order to implement the TEBD numerical scheme for the simulation of S/B coupling model, we transform the Hamiltonian by mapping the bath operators to the operators of a semi-infinite lattice chain, through a unitary transformation: $ b_n = \int_{-{x_m}}^{x_m} U_n(x)d(x) \mathrm{d}x $. In this case, we choose normalized shifted Legendre polynomial as the unitary operator $U_n(x) = \sqrt{\frac{(2n+1)}{2 x_{m}}} L_n(x/x_{m}) $ defined in the range of $x\in [-x_{m},x_{m}]$ for having spectral density in the form of Eq. \eqref{spctral_density}. The unitary operator satisfies orthogonality condition, and we obtain the transformed Hamiltonian as

\begin{align}\label{chain_ham}
	{H}_{chain} & = H_S+ \eta' \left(a^\dagger b_0+ a b_0^\dagger\right)\\
	&+\lim_{N\to \infty} \left[ \sum_{n=0}^N \omega_n b_n^\dagger b_n+ \sum_{n=0}^{N-1} \eta_n \left(b_n^\dagger b_{n+1} + b_n b_{n+1}^\dagger\right)\right]\nonumber
\end{align}

where the coefficients are $\eta' = c_0\sqrt{2 \omega_c},\omega_n = 0$ and $,\eta_n = \omega_c \left(\frac{n+1}{\sqrt{(2n+1)(2n+3)}}\right)$. Fig. \ref{fig:tebd}(a) presents a schematic diagram of the transformation. Such mappings have been used recently in Ref. \cite{openquanta_1dchain} for the simulation of open quantum systems to apply on spin-boson models \cite{openquanta_1dchain_spin_bosson} and biomolecular aggregates \cite{openquanta_1dchain_biomolecul}.

The next step is the simulation of the chain using TEBD algorithm. For that, we express the state of the chain as a matrix product state (MPS):

\begin{align}
	|\Psi \rangle = &\sum_{{\alpha _{1},.,\alpha _{N +1}=0} }^{\chi} \sum_{ i_1...i_N=0}^{M} \lambda _{\alpha _{1}}^{[1]}\Gamma _{\alpha _{1}\alpha _{2}}^{[1]i_{2}} \lambda _{\alpha _{2}}^{[2]}\Gamma _{\alpha _{2}\alpha _{3}}^{[2]i_{3}}\cdot.\\
	&..\cdot \lambda _{\alpha _{N}}^{[N]}\Gamma _{\alpha _{N}}^{[N]i_{N}} \lambda _{\alpha _{N+1}}^{[N+1]}|{i_{1},i_{2},..,i_{N-1},i_{N}}\rangle\nonumber
\end{align}

The MPS state is obtained through the Schmidt decomposition of the pure state of N sites where $\chi$ is the Schmidt number and M is the dimension of local Hilbert space. 
The method of numerical simulation for the real-time evolution is shown diagrammatically in Fig. \ref{fig:tebd} (b), where we used 2nd order Suzuki Trotter (ST) expansion which presents the unitary evolution operator as

\begin{equation}
	U_{\delta t}= e^{-i\delta tH_{chain}}=e^{-iF\delta t/2}e^{-iG\delta t}e^{-iF\delta t/2}+O[\delta t^3]
\end{equation}

where, $ F=\sum_{i\ \mathrm{odd}}H_{chain}^{i,i+1}$ and $G=\sum_{i\ \mathrm{even}}H_{chain}^{i,i+1}$. The ST expansion evolves the pairs of alternate sites, minimizing the error in 3rd order of the time step by evolving the pairs of alternate sites.

The simulation parameters are typically estimated by looking at error appearing in two ways; modeling the S/B coupling Hamiltonian to a 1D chain and each step of simulation of the real-time evolution. Previously, the extensive discussion of error and the estimation of parameters are discussed in Ref.  \cite{Agasti_kerr, Agasti_OQS}.

\begin{figure}[t]
	\centering
	\includegraphics[width=\linewidth]{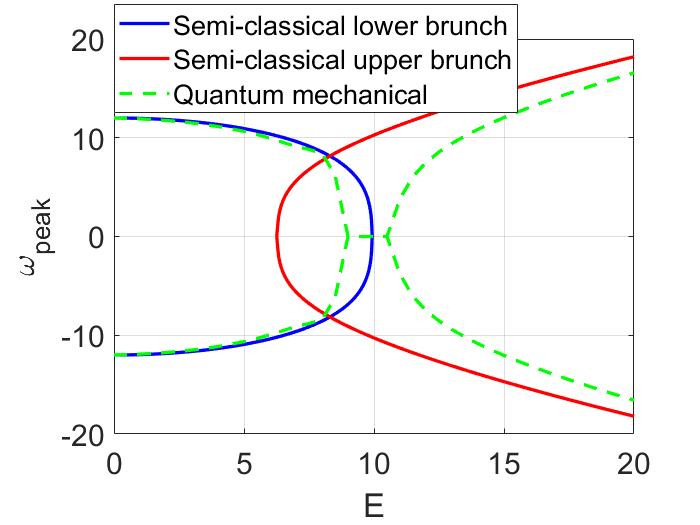}
	\caption{Normal mode splitting of the fluctuation spectra in different regime}
	\label{fig:w_split}
\end{figure}

\section{APPENDIX: SEMI-CLASSICAL EQUATION} \label{SEMI_CLASSICAL_EQUATION}	

The form the system Hamiltonian in Eq. \eqref{kerr_hamiltonian} provides a nonlinear Heisenberg equation of motion (Eq. \eqref{QLEfrKerr}) which in turn determines the semi-classical stationary value of the system field. The presence of strong coherent pump helps us to split the system field by its steady-state classical field $(\alpha\equiv\langle a\rangle_{t\infty})$ and the fluctuations around it: $a_{old} = \alpha+a_{new}$, where $\alpha$ is the steady-state system field. Neglecting the fluctuation terms, the zeroth order of the Eq. \eqref{QLEfrKerr} determines the steady-state system field given in Eq. \eqref{kerr coherent solution}.

\begin{align}\label{kerr coherent solution}
	|E|^2=|\alpha|^2\left((\Delta+2\chi"|\alpha|^2)^2+\frac{\gamma^2}{4}\right),
\end{align} 

The extreme values of lower and upper branches are determined by taking derivative over the Eq. \eqref{kerr coherent solution} $\left( \frac{\partial E^2}{\partial |\alpha|^2} =0 \right)$. The highest value of lower branch is

\begin{subequations}\label{kerr turning points}
	\begin{align}
		|\alpha|_l^2=\frac{1}{6\chi"}\left(-2\Delta -\sqrt{\Delta^2-3\frac{\gamma^2}{4}}\right) ,
	\end{align}
	\text{and the lowest value of upper branch is} 
	\begin{align}
		|\alpha|_u^2=\frac{1}{6\chi"}\left(-2\Delta +\sqrt{\Delta^2-3\frac{\gamma^2}{4}}\right) .
	\end{align}
\end{subequations}

These extreme values in turn are known as the turning points, which fixes the region for the quantum jump of the steady-state system field. This also fixes the condition to visualize bistability which is the detuning must exceed a critical value: $\Delta^2 > 3\frac{\gamma^2}{4}$.
The changes of fluctuation around the steady-state value are determined from the first order of the Eq. \eqref{QLEfrKerr}

\begin{align}\label{HEM Kerr}
	\dot{a}&=-i\Delta a -\frac{\gamma}{2}a -2i\chi" \alpha^2 a^\dagger -4i\chi" |\alpha|^2 a 
\end{align}

Notice that the nonlinear dissipative terms $-4i\chi" |\alpha|^2 a$ in this equation regulate the resonance frequency of the linearized response of the cavity field fluctuations.

\section{APPENDIX: LINEARIZED MODEL OF SYSTEM FIELD FLUCTUATIONS} \label{LINEARIZED_MODEL}

The fluctuation spectra of the system are defined as

	\begin{align}
		S_\omega^\theta &= \int_{-\infty}^\infty \mathrm{d}\tau e^{i\omega\tau} \langle\lbrace X^\theta(t+\tau),X^\theta(t) \rbrace\rangle \nonumber\\
		&=\int_{-\infty}^\infty \mathrm{d}\tau e^{i\omega\tau} \langle\lbrace (a^\dagger(t+\tau)e^{-i\theta} + a(t+\tau)e^{i\theta}), (a^\dagger(t)e^{-i\theta} + a(t)e^{i\theta}) \rbrace\rangle \nonumber \\
		&= 1+\phi(\omega)+\phi(-\omega) +\chi^\theta(\omega)
	\end{align}

where

	\begin{align}
		\phi(\omega) &= \int_{-\infty}^\infty \mathrm{d}\tau e^{i\omega\tau} \left[\langle a^\dagger(t+\tau) a(t) \rangle + \langle a^\dagger(t) a(t+ \tau) \rangle \right] = \phi(-\omega)\\
		\chi^\theta(\omega)  &= e^{-2i\theta} \int_{-\infty}^\infty \mathrm{d}\tau e^{i\omega\tau}\left[\langle a(t+\tau) a(t) \rangle + \langle a^\dagger(t) a^\dagger(t+ \tau) \rangle \right]  +c.c
	\end{align}

In the presence of a strong coherent tone, the dynamics of the system are affected by the fluctuations introduced by the nonlinear Hamiltonian. We determine the fluctuation spectra analytically from the linearized Heisenberg EOM over nonlinear stationary field amplitude. 

\begin{equation}
	\label{eq:S11}
	S_\omega^\theta = \frac{1}{2} \langle \left\{X_\omega^\theta , X_{-\omega}^\theta \right\} \rangle ,
\end{equation}
with
$X_\omega^\theta = 1/\sqrt{2} \left(a_{-\omega}^\dagger e^{i \theta} + a_\omega
e^{-i \theta}\right)$, can be obtained  from the Fourier transformation of the Eq. \eqref{HEM Kerr} and its Hermitian conjugate:
\begin{subequations}
	\begin{align}
		&\left[-i \omega + i \Delta + \frac{\gamma}{2} + 4i\chi" |\alpha|^2 \right] a_\omega + 2i\chi" \alpha^2 a_{-\omega}^\dagger = \sqrt{\gamma} a_{\mathrm{in},\omega}  \label{eq:S6} \\
		&\left[-i \omega - i \Delta + \frac{\gamma}{2} - 4i\chi" |\alpha|^2 \right] a_{-\omega}^\dagger - 2i\chi" {\alpha^*}^2 a_{\omega} = \sqrt{\gamma} a_{\mathrm{in},-\omega}^\dagger \label{eq:S7}
	\end{align}
\end{subequations}
with the usual definition of the Fourier transform:

\begin{align*}
	a_\omega = \int_{-\infty}^\infty \mathrm{d}t e^{i\omega t}a(t), \qquad \qquad a_{-\omega}^{\dagger} = \int_{-\infty}^\infty \mathrm{d}t e^{i\omega t}a^{\dagger}(t)
\end{align*}

where $a_{in}$ is the stochastic input field. Defining

\begin{subequations}
	\label{eq:S8a}
	\begin{align}
		A &= -i \omega + i \Delta + \frac{\gamma}{2} + 4i\chi" |\alpha|^2, \\
		B &= 2i\chi" \alpha^2, \\
		C &= -i \omega - i \Delta + \frac{\gamma}{2} - 4i\chi" |\alpha|^2 ,
	\end{align}
\end{subequations}

the QLE of the system can be expressed as
\begin{equation}
	\label{eq:S8b}
	\begin{pmatrix}
		a_\omega \\ a_{-\omega}^\dagger
	\end{pmatrix}
	= \frac{1}{AC - \left|B\right|^2}
	\begin{pmatrix}
		C & -B\\
		-B^* & A
	\end{pmatrix}
	\begin{pmatrix}
		\sqrt{\gamma} a_{\mathrm{in},\omega} \\
		\sqrt{\gamma} a_{\mathrm{in},-\omega}^\dagger  
	\end{pmatrix} 
\end{equation}
where $A(-\omega) = C^*(\omega)$. This gives
\begin{subequations}
	\begin{align}
		a_\omega &= \chi_\mathrm{d}\left(\omega\right) a_{\mathrm{in},\omega} + \chi_\mathrm{x}\left(\omega\right) a_{\mathrm{in},-\omega}^\dagger  , \label{eq:S9}\\
		a_{-\omega}^\dagger &= \chi_\mathrm{x}^*\left(-\omega\right) a_{\mathrm{in},\omega} + \chi_\mathrm{d}^*\left(-\omega\right) a_{\mathrm{in},-\omega}^\dagger \label{eq:S10}
	\end{align}
\end{subequations}
where
\begin{subequations}
	\begin{align}
		\chi_\mathrm{d}\left(\omega\right) &= \sqrt{\gamma} C (AC -\left|B\right|^2)^{-1}, \\ \chi_\mathrm{x}\left(\omega\right) &= -\sqrt{\gamma} B (AC -\left|B\right|^2)^{-1},
	\end{align}
\end{subequations}
As the system is coupled to an empty bath, the anticommutator of the operators of input field gives a delta function: 
$\lbrace a_{\mathrm{in},\omega} , a_{\mathrm{in},\omega '}^\dagger \rbrace =
\delta \left(\omega - \omega '\right)$. The fluctuation spectrum of the system is
%

	\begin{equation}\label{Eq: squeezing spectra}
		\begin{split}
			S_\omega^\theta =&\, \frac{1}{4} \left[ \left( \left|\chi_\mathrm{d}\left(\omega\right)\right|^2 + \left|\chi_\mathrm{x}\left(-\omega\right)\right|^2 \right) \lbrace \left\{a_{\mathrm{in},\omega}, a_{\mathrm{in},\omega}^\dagger\right\}\rbrace + \left( \left|\chi_\mathrm{d}\left(-\omega\right)\right|^2 + \left|\chi_\mathrm{x}\left(\omega\right)\right|^2 \right) \lbrace\left\{a_{\mathrm{in},-\omega}^\dagger, a_{\mathrm{in},-\omega}\right\}\rbrace \right] \\
			&+ \frac{1}{4} \Big[ \left(\chi_{\mathrm{d}}\left(\omega\right) \chi_{\mathrm{x}}\left(-\omega\right) e^{-i2\theta} + \chi_{\mathrm{d}}^* \left(\omega\right) \chi_{\mathrm{x}}^* \left(-\omega\right) e^{i2\theta}\right) \lbrace\left\{a_{\mathrm{in},\omega}, a_{\mathrm{in},\omega}^\dagger\right\}\rbrace \\
			&\qquad + \left(\chi_{\mathrm{d}}\left(-\omega\right) \chi_{\mathrm{x}}\left(\omega\right) e^{-i2\theta} + \chi_{\mathrm{d}}^* \left(-\omega\right) \chi_{\mathrm{x}}^* \left(\omega\right) e^{i2\theta} \right) \{\left\{a_{\mathrm{in},-\omega}^\dagger, a_{\mathrm{in},-\omega}\right\}\} \Big] \\
			=&\, \frac{1}{4} \Big[ \left|\chi_{\mathrm{d}}\left(\omega\right)\right|^2 + \left|\chi_{\mathrm{d}}\left(-\omega\right)\right|^2 + \left|\chi_{\mathrm{x}}\left(\omega\right)\right|^2 + \left|\chi_{\mathrm{x}}\left(-\omega\right)\right|^2  + 2 \cos\left(2\theta - \phi\right) \left| \chi_{\mathrm{d}}\left(\omega\right) \chi_{\mathrm{x}}\left(-\omega\right) + \chi_{\mathrm{d}} \left(-\omega\right) \chi_{\mathrm{x}} \left(\omega\right) \right| \Big] 
		\end{split}
	\end{equation}

%
where
$\phi = \mathrm{Arg} \left[ \chi_{\mathrm{d}}
\left(\omega\right) \chi_{\mathrm{x}} \left(-\omega\right) +
\chi_{\mathrm{d}} \left(-\omega\right)
\chi_{\mathrm{x}} \left(\omega\right) \right]$. 

The denominator of $S^\theta_\omega$ is 

	\begin{align*}
		|AC-|B|^2|^2 = |(-i \omega + i \Delta + \frac{\gamma}{2} + 4i\chi" |\alpha|^2 )(-i \omega - i \Delta + \frac{\gamma}{2} - 4i\chi" |\alpha|^2) - (2 \chi" |\alpha|^2)^2|^2
	\end{align*}

The poles of $S^\theta_\omega$ are located at 

\begin{equation}\label{split freq}
	\omega = \frac{1}{2}  \left[\pm i\gamma \mp 2 \sqrt{(\Delta +4\chi" |\alpha|^2)^2 - 4 \chi"^2 |\alpha|^4} \right]
\end{equation}

We see normal mode splitting in the cavity field amplitude introduced when the condition

\begin{align*}
	(\Delta +4\chi" |\alpha|^2)^2 \ge 4 \chi"^2 |\alpha|^4
\end{align*}

satisfies, which simplifies to

\begin{subequations}
	\begin{align}
		|\alpha|_{lw}^2 \le -\Delta/6\chi"\\
		|\alpha|_{up}^2 \ge -\Delta/2\chi"
	\end{align}
\end{subequations}

$|\alpha|_{up}$ and $|\alpha|_{lw}$ are the extreme limits belonging to the upper and lower branches, respectively.

The normal mode splitting is plotted in Fig. \ref{fig:w_split} for both the semi-classical and quantum mechanical estimations.
It is worth noticing that the splitting of frequency modes starts within the boundary of turning points:

\begin{subequations} 
	\begin{align}
		|\alpha|_l^2=\frac{1}{6\chi"}\left(-2\Delta -\sqrt{\Delta^2-3\frac{\gamma^2}{4}}\right) \ge -\Delta/6\chi" ,\\
		|\alpha|_u^2=\frac{1}{6\chi"}\left(-2\Delta +\sqrt{\Delta^2-3\frac{\gamma^2}{4}}\right) \le -\Delta/2\chi".
	\end{align}
\end{subequations}

Therefore, the semi-classical turning points do not satisfy the necessary conditions for normal mode splitting. That is why we don’t see any normal mode splitting at turning points in Fig. \ref{fig:w_split}. In the lower branch, the splitting decreases when the steady-state field amplitude increases with the increment of the driving field. On the other side, the splitting increases in the upper branch with the increment of the steady-state field amplitude.

\end{document}